\documentclass[reprint,superscriptaddress,amsmath,amssymb,aps]{revtex4-2}
\usepackage{graphicx}
\usepackage{dcolumn}
\usepackage{bm}
\usepackage{siunitx} 

\usepackage{pdfpages} 
\usepackage{pgffor} 

\makeatletter
\AtBeginDocument{\let\LS@rot\@undefined}
\makeatother

\def\supplementfilename{supplemental.pdf}
    
\pdfximage{\supplementfilename}
\def\numbersupplementpages{\the\pdflastximagepages}

\def\l{{\langle}}
\def\r{{\rangle}}
\def\n{{\rm n}}
\def\x{{\mathbf{x}}}
\begin{document}
 
\title{Nonlinear Diffusion and Decay of an Expanding Turbulent Blob}

\author{Takumi Matsuzawa}
\thanks{These authors contributed equally}
\affiliation{James Franck Institute and Department of Physics, University of Chicago, Chicago, IL 60637, USA}
\affiliation{Present address: Laboratory of Atomic and Solid-State Physics, Cornell University, Ithaca, NY, USA}

\author{Minhui Zhu}
\thanks{These authors contributed equally}
\affiliation{Department of Physics, University of Illinois at Urbana-Champaign, Loomis Laboratory of Physics, 1110 West Green Street, Urbana, Illinois 61801-3080, USA}
\affiliation{Data Science and Learning Division, Argonne National Laboratory, Lemont, Illinois 60439, USA}

\author{Nigel Goldenfeld}
\affiliation{Department of Physics, University of California, San Diego, 9500 Gilman Drive, La Jolla, California 92093, USA}

\author{William T.M. Irvine}
\affiliation{James Franck Institute and Department of Physics, University of Chicago, Chicago, IL 60637, USA}
\affiliation{Enrico Fermi Institute, University of Chicago, Chicago, IL 60637, USA}

\date{\today}

\begin{abstract}
Turbulence, left unforced, decays and invades the surrounding quiescent fluid. Though ubiquitous, this simple phenomenon has proven hard to capture within a simple and general framework. 
Experiments in conventional turbulent flow chambers are inevitably complicated by proximity to boundaries and mean flow,  obscuring the fundamental aspects of the relaxation to the quiescent fluid state. 
Here, we circumvent these issues by creating a spatially-localized blob of turbulent fluid using eight converging vortex generators focused towards the center of a tank of water, and observe its decay and spread over decades in time, using particle image velocimetry with a logarithmic sampling rate. 
The blob initially expands and decays until it reaches the walls of the tank and eventually transitions to a second regime of approximately spatially uniform decay. 
We interpret these dynamics within the framework of a nonlinear diffusion equation, which predicts that the ideal quiescent-turbulent fluid boundary is sharp and propagates non-diffusively, driven by turbulent eddies while decaying with characteristic scaling laws. 
We find direct evidence for this model within the expansion phase of our turbulent blob and use it to account for the detailed behavior we observe, in contrast to earlier studies. 
Our work provides a detailed spatially-resolved narrative for the behavior of turbulence once the forcing is removed, and demonstrates unexpectedly that the turbulent cascade leaves an indelible footprint far into the decay process.
\end{abstract}

\maketitle

The energy cascade is arguably the most distinctive feature of three dimensional (3D) steady state turbulence, setting it apart from other forms of random dynamics such as chaos. In the turbulent cascade, energy is transferred from the largest scale $\ell$ at which the flow is forced, via a Hamiltonian process to viscous scales, where it is ultimately dissipated. Typically the turbulent cascade is generated from an imposed mean flow, as in a pipe or wind-tunnel, but many other flow geometries have been explored, with the consensus that regardless of the large scale generation, the small-scale behavior is universal, even at intermediate levels of turbulence \cite{schumacher2014small,cerbus2020small}, and characterized by a suite of scaling laws for the distribution of energy across scales that have been the focus of theoretical efforts to understand the statistical properties of turbulence \cite{kolmogorovLocalStructureTurbulence1941,kolmogorovDissipationEnergyLocally1941,kolmogorovRefinementPreviousHypotheses1962,popeTurbulentFlows2000,sreenivasan2025turbulence}.  Turbulence has also been extensively explored in quantum systems (for an introductory review, see \cite{tsubota2013quantum,barenghi2014introduction,barenghi2023quantum}), including experimentally in the regime where normal and superfluid components are coupled \cite{zhang2023higher}, thus emulating classical turbulence in both decay \cite{smithDecayVorticityHomogeneous1993,skrbekFourRegimesDecaying2000} and spectral properties \cite{zhao2025kolmogorov}.

If the forcing stops, this cascade is starved, and turbulence decays, ultimately entering the laminar regime.
Although there is still not a consensus on a satisfactory theoretical framework for the decay,  a variety of experimental set-ups have been applied to study the decay process, often in the context of a superimposed mean flow where the decay in time is transformed to a distance-dependent scaling.  Examples include passive grids~\cite{hurstScalingsDecayFractalgenerated2007, kistlerGridTurbulenceLarge1966, comte-bellotSimpleEulerianTime1971a, bodenschatzVariableDensityTurbulence2014}, active grids~\cite{hideharuRealizationLargescaleTurbulence1991, mydlarskiOnsetHighReynoldsnumberGridgenerated1996}, pipes~\cite{reynoldsXXIXExperimentalInvestigation1883, mullinExperimentalStudiesTransition2011, eckhardtTurbulenceTransitionPipe2007, hofScalingTurbulenceTransition2003}, and spinning disks~\cite{delatorreSlowDynamicsTurbulent2007a, volkDynamicsInertialParticles2011, labbeStudyKarmanFlow1996}, with reported putative power laws for energy decay spread over a large range~\cite{desilvaOscillatingGridsSource1994, georgeDecayHomogeneousIsotropic1992, sinhuberDecayTurbulenceHigh2015, panickacheriljohnLawsTurbulenceDecay2022,gorce2024freely}, often deviating from the necessarily {\it ad hoc} theoretical predictions~\cite{dekarmanStatisticalTheoryIsotropic1938, batchelorDecayIsotropicTurbulence1948, batchelorDecayTurbulenceFinal1948, batchelorTheoryHomogeneousTurbulence1953, saffmanNoteDecayHomogeneous1967}.
Even under idealized conditions of spatial homogeneity, theoretical and experimental results for decay rates span a broad spectrum~\cite{smithDecayVorticityHomogeneous1993,smithStudyHomogeneousTurbulence1994,skrbekDecayHomogeneousIsotropic2000, sinhuberDecayTurbulenceHigh2015, panickacheriljohnLawsTurbulenceDecay2022, kuchlerUniversalVelocityStatistics2023, georgeDecayHomogeneousIsotropic1992, davidsonTurbulenceIntroductionScientists2015,gorce2024freely}.

Explorations of the decay of turbulence are confounded by difficulties such as finite Reynolds number, finite system size, presence of mean flows, and interactions with boundaries. Furthermore, in natural settings, turbulence is rarely homogeneous. Instead, it often exhibits fronts, with turbulent regions interfacing quiescent zones across distinct boundaries characterized by abrupt changes in vorticity and Reynolds stresses ~\cite{dasilvaIntenseVorticityStructures2011}.
In such settings, the free evolution of turbulence inherently involves both decay and propagation into surrounding quiescent regions. No unified body of experiment or theory currently integrates these intertwined processes of turbulent transport and decay in a regime where the laminar-turbulent front dynamics can be systematically probed.

Here, we introduce a new experimental platform for probing the decay and propagation of turbulence unconstrained by boundaries or mean flows. We report compelling evidence for the nonlinear diffusion of the laminar-turbulent fronts as a blob of turbulence expands into a laminar region of fluid, and then subsequently transitions into a homogeneous decay regime at long times, all of which are consistent with predictions based upon a phenomenological model for inhomogeneous turbulence \cite{kolmogorov1942equations, barenblatt1983self, barenblattEvolutionTurbulentBurst1987,chen1992renormalization}.

Our experimental platform for investigating the free evolution of turbulence uses vortex rings to produce an isolated region of turbulence in the middle of a tank~\cite{matsuzawaCreationIsolatedTurbulent2023}.
By localizing turbulence at the center of the chamber and minimizing mean flow and boundary effects, our setup enables direct observation of the intrinsic behaviors of turbulence as it propagates, mixes with the surrounding fluid, and decays toward thermal equilibrium.

\begin{figure}[htbp]
\centering
\includegraphics[width=0.9\columnwidth]{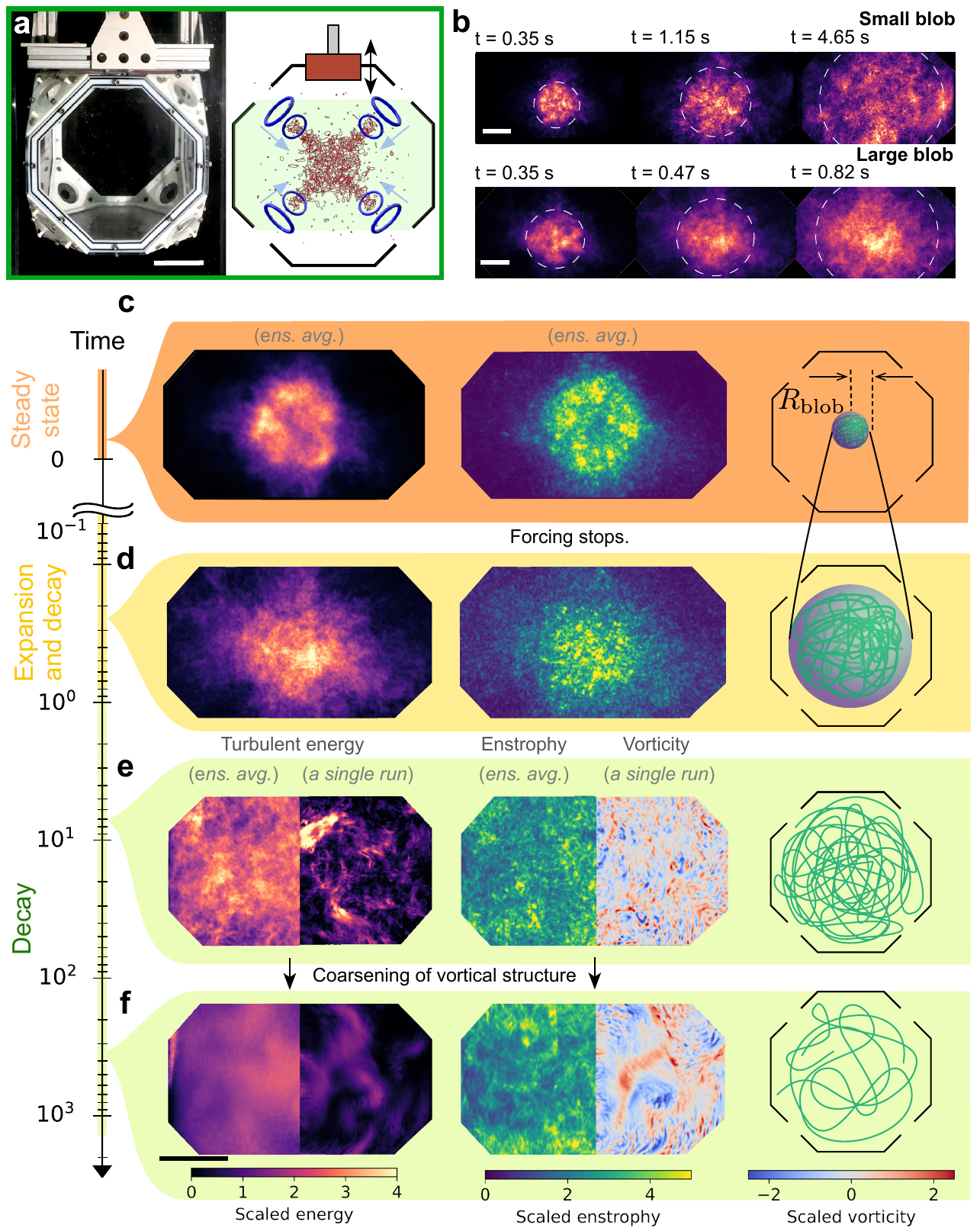}
\caption{\textbf{Propagation and decay of confined turbulence (large blob, Re$_\lambda=203$).} (\textbf{a}) Left: Experimental setup of a vortex ring collider. Right: A schematic of the experiment is shown. We repeatedly fire vortex loops toward the center of the chamber to create an isolated blob of turbulence. (\textbf{b}) Once vortex rings are no longer supplied, the turbulent blobs expand, as visualized by turbulent energy.  Small and large blob configurations display distinct rates of expansion (Small: Re$_\lambda=60$, $\n=10$,  Large: Re$_\lambda=203$, $\n=21$).
(\textbf{c}) The images display the ensemble-averaged, fluctuating energy and enstrophy fields in the state of a turbulent blob (Large blob). (\textbf{d}) The ensemble-averaged fluctuating energy and enstrophy fields show the expansion of the blob shortly after the forcing stops. (\textbf{e}) Once the expansion is complete, the ensemble-averaged fields display that turbulence is homogeneously distributed as illustrated on the left side of each panel. The right side shows a single experimental run, and reveals swirling motions across the scales both in energy and vorticity fields. 
(\textbf{f}) As turbulence continues to decay, the vortical structure coarsens over time until the flow reaches a thermal equilibrium. The scale bars in (a) and (b) represent 100 mm and 50 mm, respectively.}
\label{fig: fig1}
\end{figure}

Figure~\ref{fig: fig1}a shows our experimental chamber.
Pulling the piston upward produces sets of eight vortex rings that travel toward the center of the tank and collide, feeding energy and enstrophy into the turbulent blob at the center.
Inside the blob, the flow is dominated by turbulent fluctuations, with fluctuating energy exceeding the energy of the time-averaged mean flow by up to a factor of 1000~\cite{matsuzawaCreationIsolatedTurbulent2023}. 
Crucially, the tunable properties of injected vortex rings directly determine the intensity and length scales of the resulting isolated blob of turbulence, as well as other inviscid invariants of the turbulent state such as angular impulse and helicity. 

When the turbulent blob is suddenly starved of incoming vortex rings, it begins to expand and decay. 
In a typical experiment, the characteristic velocities drop from approximately 150 mm/s just before the forcing stops to 0.4 mm/s at the end of the decay. This significant dynamic range and long decay duration pose a challenge for recording Particle Image Velocimetry (PIV) data, which requires tracer particles to move at least about one diameter between frames for accurate measurements of velocity vectors~\cite{raffelParticleImageVelocimetry2018a}.
We overcome this challenge by continuously varying the spacing between recorded frames from 2.5 ms to 1 s linearly (See SM \S 3A for the data acquisition protocol). 
This approach generates a movie with frames spaced evenly in logarithmic time, rather than at uniform intervals in linear time like conventional movies (see Supplementary Movie 8).

Our ``log-movie'' approach is broadly applicable to self-similar phenomena governed by power-law scaling which spans many orders of magnitude, from gravity currents~\cite{huppert2006gravity} to phase separation~\cite{gaulin1987kinetics, khain2008generalized}.
For the turbulent blob, this method enables us to capture PIV data over a 17-minute span, during which the kinetic energy at the tank's center decreases by approximately a factor of $10^5$  (see Supplementary Movie 4). 
By the end of the recording, the flow appears quiescent to the naked eye.
In a later section, we show that despite the apparent stillness, the flow retains distinct turbulent statistics even at $\mathrm{Re}_\lambda \approx 10$.

We repeat our protocol 21 times to construct an {\it ensemble of decays}. 
Using this ensemble, we perform a Reynolds decomposition, separating the total velocity field $U_i({\bf{x}}, t)$ into a time-dependent mean flow $\langle U_i \rangle_{\rm n} ({\bf{x}}, t)$ and a fluctuating velocity field $u_i({\bf{x}}, t) = U_i({\bf{x}}, t) - \langle U_i \rangle_{\rm n} ({\bf{x}}, t)$, as shown in Figure~\ref{fig: fig1}b.
Here, the subscript $\rm{n}$ represents an ensemble average.
While individual realizations display anisotropic expansion of energy and vorticity, ensemble-averaging reveals a confined region of initially homogeneous isotropic turbulence (HIT)~\cite{matsuzawaCreationIsolatedTurbulent2023} that decays and expands simultaneously, progressively filling the flow chamber (Figure~\ref{fig: fig1}c–d and Supplementary Movies 4 and 6).
The late-time decay, marked by the coarsening of turbulent whorls, is vividly depicted in Figure~\ref{fig: fig1}f.

To estimate the total turbulent energy within the system, we begin by spatially averaging the turbulent energy field, $q({\bf{x}}, t)= u_i u_i /2$, over the 2D PIV slice for each recording, and then ensemble-average over n recordings. 
The resulting quantity, $\langle q \rangle _{\mathbf{x}, \rm{n}}(t)$, represents the average energy density—turbulent energy per unit volume in the chamber. 
In homogeneous isotropic turbulence, this quantity is proportional to the total turbulent energy in the chamber. This assumption holds throughout the grid experiment and during the final decay phase of the blob experiment, when the turbulent energy distribution becomes nearly uniform. However, it breaks down during the blob’s expansion, when the energy field is strongly inhomogeneous.

\begin{figure}[htbp]
\centering
\includegraphics[width=0.85\columnwidth]{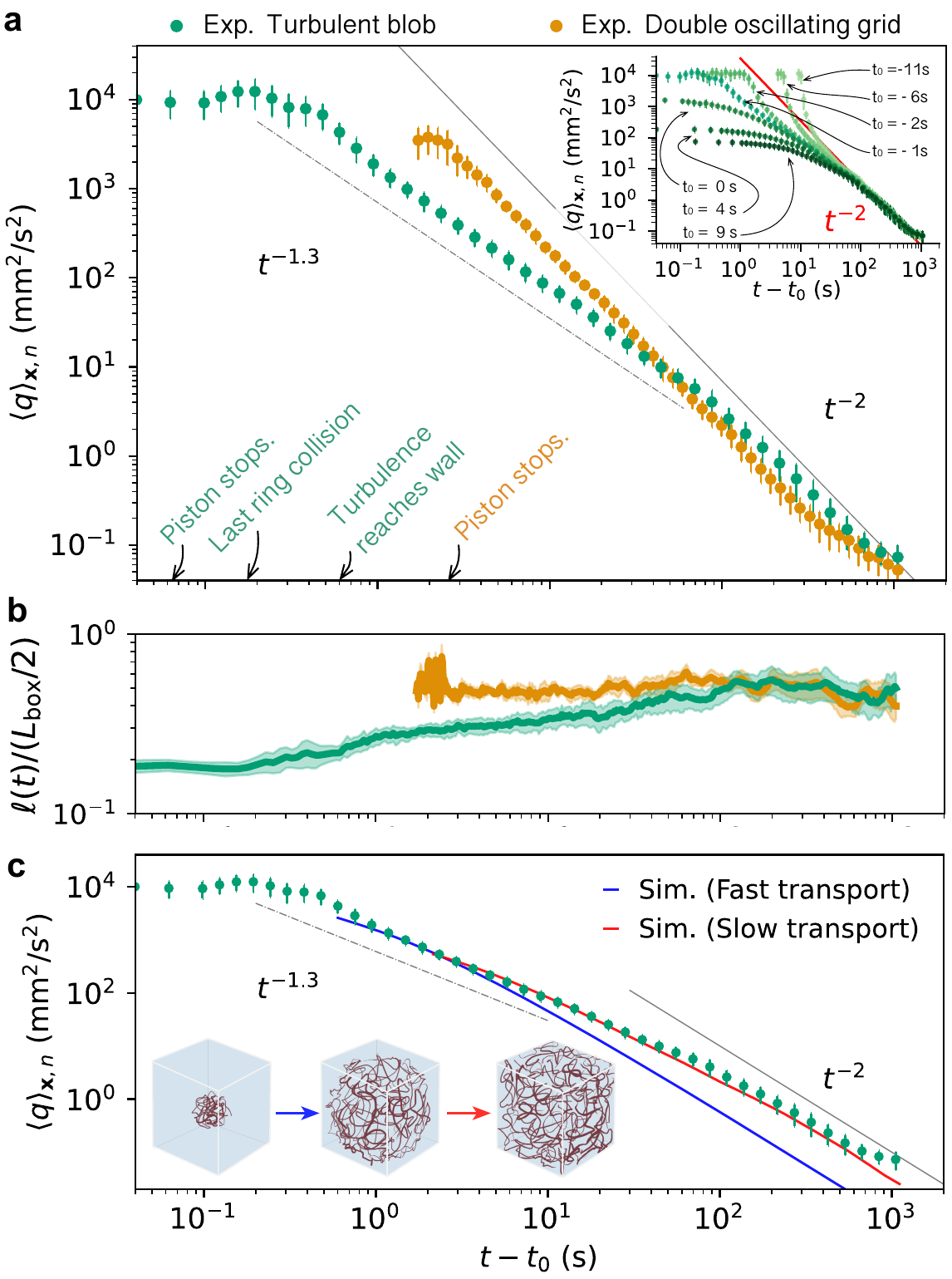}
\caption{\textbf{Growth of integral lengths scale dictates the decay exponent of turbulent kinetic energy.}  (\textbf{a}) Distinct decay behaviors of turbulent kinetic energy density are observed for two systems: a large turbulent blob (Re$_\lambda=203$, $\n=21$) and a double oscillating grid ($\n=10$). The turbulent blob exhibits a dual decay regime, characterized by the decay exponents of $-1.3$ and $-2$, contrasting with the uniform power-law decay with an exponent of $-2$ in the double oscillating grid setup, with virtual origins at $-0.15$ s and $-2.72$ s, respectively. (\textbf{b}) The integral length scale, $\ell(t)$, exhibits a power-law growth $(t-t_0)^{0.16}$, starting from approximately the radius of the blob until it reaches saturation. 
\textbf{(c)} Using experimentally determined initial conditions and the integral length scale, the 3D cell dynamical systems simulations reproduce the observed power laws. Lowering the transport coefficient $c_0$ delays the onset of the crossover. The coefficients for the fast/slow transport are ($c_0, \epsilon_0$)=(1.2, 0.88) and (0.001, 0.9), respectively. The fast transport uses ensemble-averaged experiment profile at $t=0.4 s$ as the initial condition; the slow one uses $t=2$ s. The error bars represent S.E.M.
}
\label{fig: fig2}
\end{figure}

The wide dynamic range of decay, along with its varying rate, motivates the use of logarithmic axes for both time and energy. As shown in the inset of Figure~\ref{fig: fig2}a, the early-time behavior of the decay curve is highly sensitive to the choice of the time origin $t_0$: different choices of $t_0$ values noticeably affect the first two-thirds of the curve. 
In the final third, the influence of $t_0$ becomes negligible and the decay settles into a robust power-law scaling with an exponent of $-2$, extending over roughly one decade.

Does the first two-thirds of the decay also follow a power law?
To answer this, we develop a virtual-origin fitting method to handle the challenges from restricted temporal range of data and the unknown time origin $t_0$.
Assuming a power-law form $(t-t_0)^m$, we scan a range of exponents $m$, plotting $\langle q(\mathbf{x}, t) \rangle_{\mathbf{x}, \rm{n}}^{1/m}$ for each; if the chosen $m$ aligns with the data, the transformed curve becomes linear and, and its x-intercept identifies the corresponding optimal $t_0$.
To assess linearity in a way that is robust to noise and independent of scale, we introduce a novel cross-validation-based $r^2$ metric (see Supplementary Information \S4C for details).
Applying this method to the first two-thirds of the data yields a best-fit exponent of $m=-1.3$ with $t_0=-0.15$~s. 
This inferred $t_0$ closely matches the arrival time of the final set of vortex rings at the blob ($t=0.2$ s), indicating the physical significance of the outcome of the fitting procedure. 
Repeating the procedure for the final third of the data again confirms a consistent power law with exponent $-2.0$.
Whether one interprets the decay as two distinct power-law regimes or not, the sharp transition in slope around $t - t_0= 80$~s stands out as a clear and robust feature of the decay.

What drives this transition in the decay mode? 
A natural hypothesis emerges if one assumes that the central turbulent region of the flow remains homogeneous and isotropic throughout the decay, with its integral scale varying slowly over time.
Under this assumption, the energy dissipation rate at small scale is governed by spatial structure of energy at large scales, instead of viscosity $\nu$ \cite{drydenReviewStatisticalTheory1943, vassilicosDissipationTurbulentFlows2015, sreenivasanScalingTurbulenceEnergy1984, davidsonTurbulenceIntroductionScientists2015}:
\begin{align}
    \epsilon (t) = \epsilon_0  \frac{\langle q^{3/2} ({\bf{x}}, t) \rangle_{\bf{x}, \rm{n}}} {\ell(t)}
    \label{eq:turb_dissipation}
\end{align}
where $\ell(t)$ is the integral length scale defined as $\ell(t) \equiv (3\pi/4)\int_0^\infty \kappa^{-1} E(\kappa, t) d\kappa~/~\int_0^\infty E(\kappa, t) d\kappa$, and $E(\kappa, t)$ is the energy spectrum. 
The evolution of $\ell(t)$, which is not prescribed by any definitive current theory,  naturally emerges as a key control parameter for the decay rate. 
Notably, in the special case where $\ell$ remains constant, the energy balance equation $d \langle q(\mathbf{x}, t) \rangle_{\mathbf{x}, \rm{n}} / dt = -\epsilon(t)$ yields a power-law decay with an exponent of $-2$, precisely matching the late-time behavior we observe.

Given the extended decay period of 17 minutes and the near-complete extinction of visible flow by the end, it is natural to ask whether the flow remains consistent with homogeneous isotropic turbulence throughout.
To address this question, we measured several key statistical indicators of HIT throughout the decay, namely: the energy spectrum, the second-order structure function, the integral length scale and the energy dissipation rate.
Figure~\ref{fig: fig4}a compares the measured decay rate of average turbulent energy, $-d \langle q(\mathbf{x}, t) \rangle_{\mathbf{x}, \rm{n}} / dt$, to the dissipation rate predicted by the HIT, $\epsilon_0 \langle q^{3/2}(\mathbf{x}, t) \rangle_{\mathbf{x}, n} / \ell(t)$.
The agreement between these two quantities is remarkably good throughout most of the experiment, deviating only near the very end as the flow approaches a quasistatic state.
Figure~\ref{fig: fig4}d and \ref{fig: fig4}f show the evolution of the energy spectrum $E(\kappa, t)$ and the second order structure function $D_{LL}(r, t)$ respectively, while the integral length scale $\ell(t)$ presented in Figure \ref{fig: fig2}b is computed from the energy spectrum.

Once the initial period has passed ($t \geq $80 s), the bare spectrum (Figure~\ref{fig: fig4}d inset) no longer exhibits the classical Kolmogorov $\kappa^{-5/3}$ scaling.
At this stage, a vortex comparable in size to the chamber persists, slowly churning the fluid.
Yet, remarkably, as shown in Figure~\ref{fig: fig4}d, when the energy spectra are rescaled by the measured dissipation rate, they collapse onto a universal profile obtained from direct numerical simulations (DNS) of homogeneous isotropic turbulence~\cite{li2008public} (see Supplementary Movie 3).
Notably, the absence of $\kappa^{-5/3}$ scaling remains consistent with Kolmogorov’s first self-similarity hypothesis~\cite{kolmogorovLocalStructureTurbulence1941}, which pertains to small-scale universality rather than to spectral form. Small-scale universality has been observed even at low Reynolds numbers—as low as $\mathrm{Re}_\lambda = 23$—in turbulent boundary layers, wakes, and grid turbulence~\cite{uberoi1969spectra, tieleman1967viscous, comte-bellotSimpleEulerianTime1971a}, as well as in turbulent puffs at $\mathrm{Re} = 1600$~\cite{cerbus2020small}.
The sudden deviation from the universal profile at high wavenumbers at later time between $t = 100 - 1000$ s is attributed to spectral leakage due to the aperiodic and truncated PIV-extracted velocity field~\cite{matsuzawaCreationIsolatedTurbulent2023}. 

The spatial second-order structure function is immune to such contamination while providing information equivalent to the energy spectrum~\cite{davidsonTurbulenceIntroductionScientists2015}. 
As shown in Figure~\ref{fig: fig4}f,  rescaling the structure functions by the dissipation rate expected for HIT also shows agreement with Kolmogorov's 2/3 power law during the initial stage of decay, followed by the shrinkage of the inertial subrange.
During the following stages of decay, the rescaled structure functions continue to align with the dissipation range of the universal curve, mirroring the behavior seen in the energy spectra.
Taken together, these observations confirm that the statistics of velocity fluctuations remain consistent with the expectations of homogeneous isotropic turbulence throughout the entire 17-minute decay.

\begin{figure}[htbp]
\centering
\includegraphics[width=0.95\columnwidth]{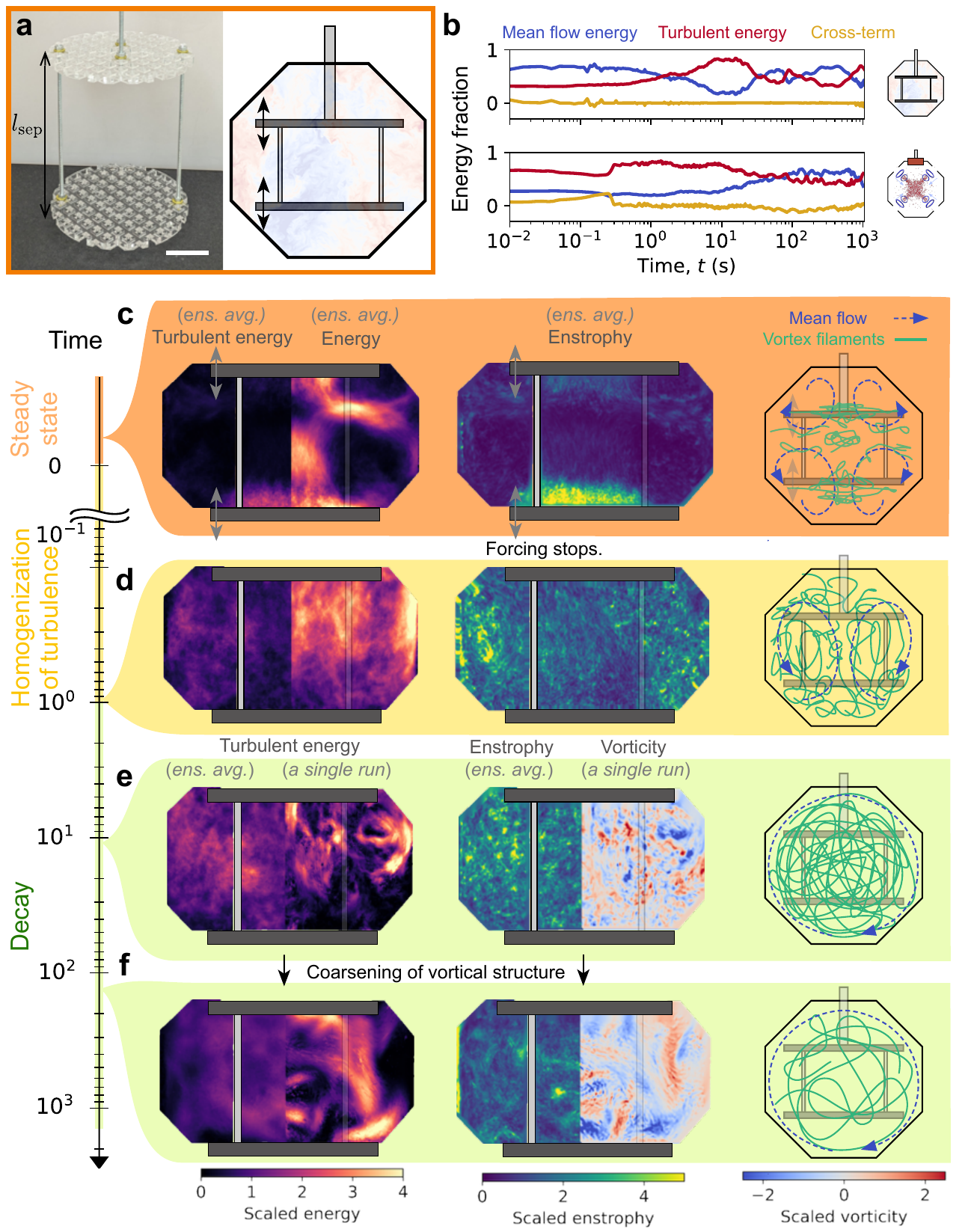}
\caption{\textbf{Temporal evolution of turbulence generated by a double oscillating grid.}  (\textbf{a})  Two grids connected by brass rods, separated by $\ell_{\rm sep} = 210$ mm, oscillate sinusoidally to agitate a fluid. The scale bar represents 50 mm. (\textbf{b}) The system's energy $\langle \textbf{U}\cdot \textbf{U}\rangle_\n$ is divided into mean flow energy $\langle \textbf{U} \rangle_\n \cdot \langle \textbf{U}\rangle_\n$, turbulent energy $\langle \textbf{u}\cdot \textbf{u}\rangle_\n$, and a cross-term $\langle \textbf{U}\cdot \textbf{u}\rangle_\n$ (blob: $\n = 21$; double oscillating grid: $\n=10$). For the double oscillating grid, the oscillation creates a substantial mean flow, whereas turbulence developing later. (\textbf{c-f}) The panels display how turbulence develops and decays. (\textbf{c}) Initially, energy is stored in a mean flow. (\textbf{d}) After the grid stops moving, turbulence develops as more mean flow energy converts into the fluctuating energy. Enstrophy becomes more homogeneous as it gets transported into the entire chamber. (\textbf{e-f}) Turbulence continues to decay, and its vortical structure coarsens over time. }
\label{fig: fig3}
\end{figure}

\begin{figure}[htbp]
\centering
\includegraphics[width=0.9\columnwidth]{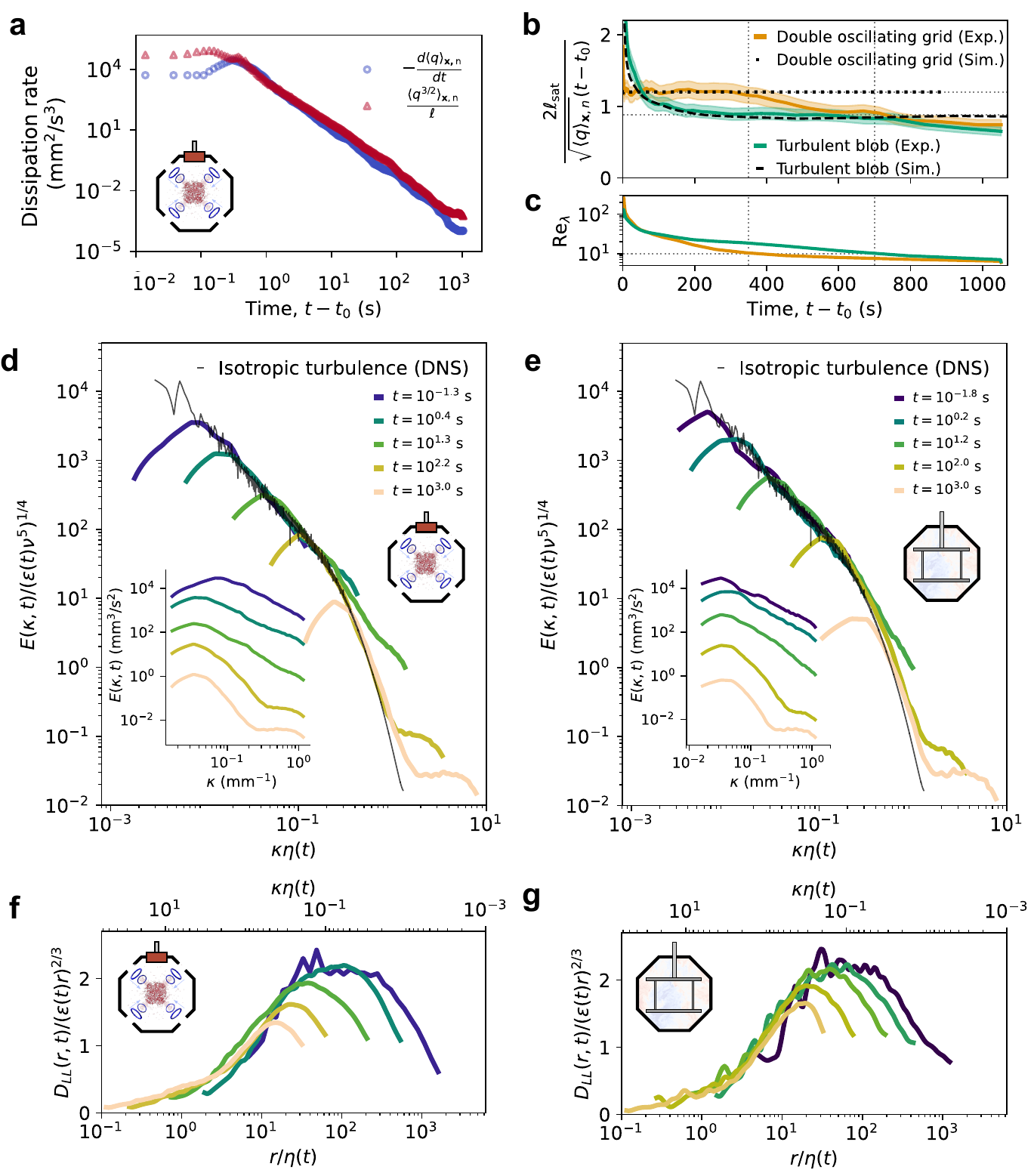}
\caption{\textbf{Enduring characteristics of turbulence through the three-dimensional energy spectrum and second-order longitudinal structure function} (\textbf{a}) The dissipation rate $-d\langle q \rangle_{\bf{x}} / dt$ for turbulence initiated by the vortex ring collisions (large blob, Re$_\lambda=203$) confirms the scaling law $\langle q^{3/2} \rangle_{\bf{x}, \n}/\ell$, consistent over an extensive duration over 500 seconds. (\textbf{b}) Estimate of dimensionless dissipation rate $\epsilon_0$ from the asymptotic form of $\l q \r_{\mathbf{x}, n}(t)$ and the saturated value of $\ell(t)$. (\textbf{c}) The Taylor Reynolds number drops below 10 when the energy decay law deviates from the $t^{-2}$. 
(\textbf{d}) Rescaled three-dimensional energy spectrum for the large turbulent blob; the inset shows spectra without rescaling, computed using the observed dissipation rate $\epsilon = -d\langle q \rangle_{\mathbf{x},\n}/dt$ and the fluid viscosity $\nu$. The rescaled spectra align with the universal profile of isotropic turbulence (DNS) at $\mathrm{Re}_\lambda = 418$ \cite{li2008public}.
(\textbf{e}) Rescaled three-dimensional energy spectrum during decay: double oscillating grid (Re$_\lambda=185$).
(\textbf{f–g}) Time evolution of the second-order longitudinal structure function during decay of a large turbulent blob (f) and turbulence generated by a double oscillating grid (g). All quantities in panels (a–g) are ensemble-averaged (large blob: $n = 21$; double oscillating grid: $n = 10$). The error in (b) represents the S.E.M.}
\label{fig: fig4}
\end{figure}

Having established the consistency with HIT throughout the decay, we now examine whether the time evolution of $\ell(t)$ is responsible for the observed sudden shift in the decay rate.
In the initial period of decay ($t - t_0 < 80$ s), the eddies grow over time as the coarsening of the vortical structure can be seen in Figure~\ref{fig: fig1}e-f and Figure~\ref{fig: fig3}e-f.
At $t-t_0 = 80$ s, when the energy decay enters the regime of $(t-t_0)^{-2}$, $\ell$ shows a clear kink at the same position. 
As shown in Figure~\ref{fig: fig2}b, our analysis confirms such growth and demonstrates that the moment $\ell$ saturates coincides with the transition of the decay law into $(t-t_0)^{-2}$. 
Accurate extraction of $\ell$ relies heavily on the energy spectrum which is prone to spectral leakage. This leakage results in underestimation of $\ell(t)$; however, we observe $\ell(t)$ eventually saturates at later times even when windowing techniques are applied to mitigate the spectral leakage.

To further test the role of $\ell(t)$ in governing the decay rate, we set out to engineer turbulence {\it in the same flow chamber} with a constant $\ell$.
To do this, we closed the orifices on the chamber's corners and replaced the acrylic piston—originally used to draw fluid through the orifices—with two acrylic grids connected by brass rods, as shown in Figure~\ref{fig: fig3}a. 
The profile of the grids, designed with a square mesh and circular outline with a solidity of 34\%, was chosen based on~\cite{desilvaOscillatingGridsSource1994} to foster the production of homogeneous turbulence. 

When the grid oscillation stops, the residual mean flow spreads vorticity throughout the chamber, initiating a transition to a more homogeneous turbulent state. Supplementary Movie 9–10 show the advection of turbulent energy and enstrophy by the residual mean flow.
Figure~\ref{fig: fig3}d confirms that the vorticity distribution is reasonably uniform.
At this stage, as shown in Figure~\ref{fig: fig3}e, the multi-scale nature of turbulence emerges, unimpeded by residual anisotropy from the initial mean flow.
Following this, the turbulence decay mirrors the same dynamics observed in the late stage of the turbulent blob, with viscosity dissipating smaller-scale motions.
As shown in Figure~\ref{fig: fig3}f, the vortical structure gradually coarsens until it reaches a single vortex state.

Similar to the turbulent blob, the energy spectrum and structure function of the fluctuating flow remain consistent with homogeneous isotropic turbulence during decay (see Supplementary Movie 12). However, the energy follows a different trend, without the kinks or abrupt changes in decay rate seen in the blob (see Supplementary Movie 9 and 11). 
Only the first few seconds of the decay are sensitive to the choice of the time origin. Applying the same fitting method for a power law, the scaling is found to be $t^{-2}$, which holds throughout the entire decay.
The ratio of mean flow to turbulent kinetic energy also evolves differently throughout the decay with the relative energy content oscillating between the two in the case of the double-grid, as shown in Figure~\ref{fig: fig3}b.
Crucially, the computation of $\ell(t)$ reveals a constant integral length scale throughout the decay. 
Thus in both cases, when $\ell(t)$ is constant, the decay of energy is consistent with $t^{-2}$,
in agreement with experiments using a towed grid~\cite{smithDecayVorticityHomogeneous1993, smithStudyHomogeneousTurbulence1994, stalpDecayGridTurbulence1999}.

\begin{figure}[htbp]
\centering
\includegraphics[width=0.90\columnwidth]{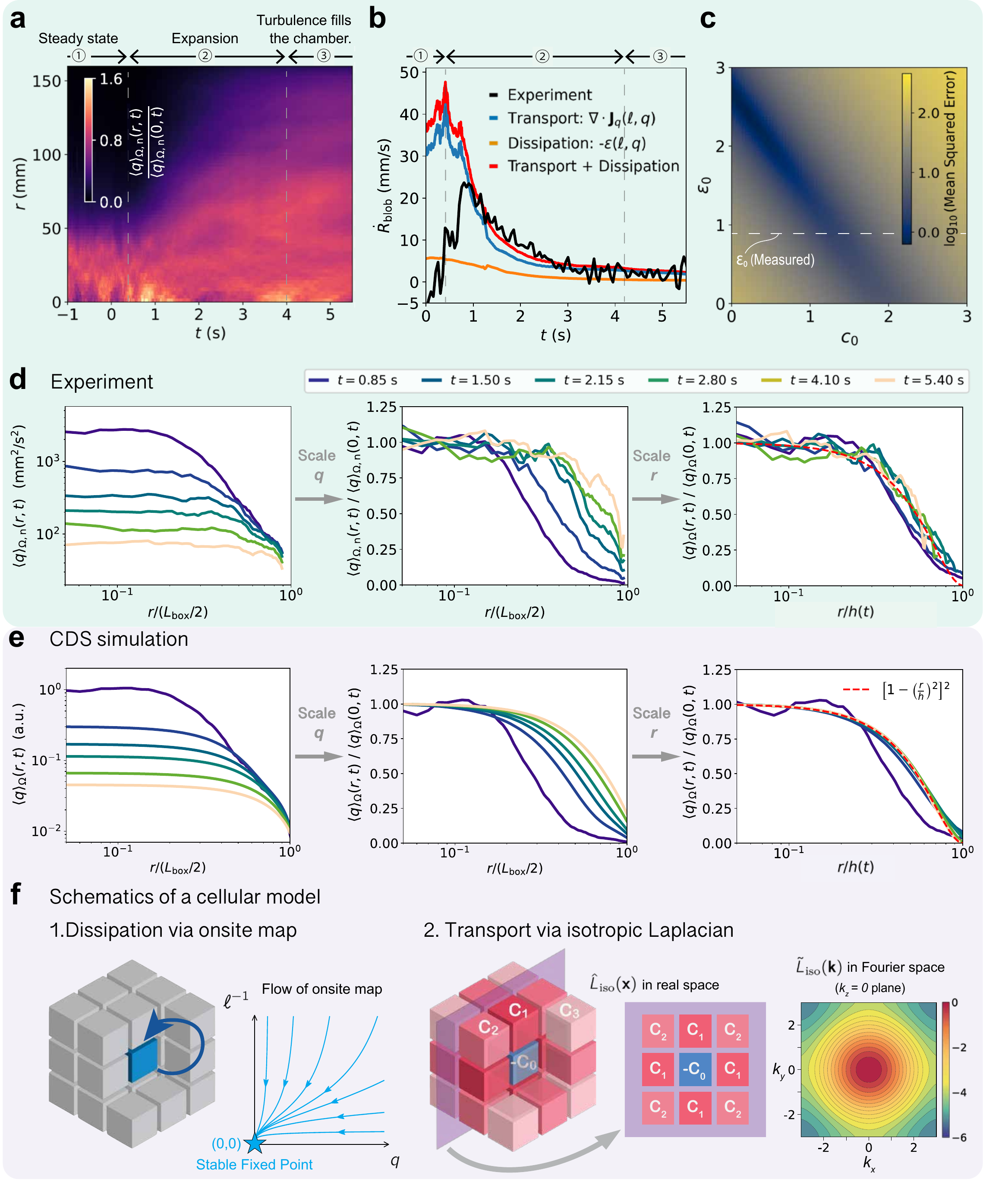}
\caption{\textbf{Front propagation and dynamic scaling during turbulence spreading.} 
(\textbf{a}) Kymograph of the turbulent energy density  $q(\mathbf{x},t)$, averaged over azimuthal angles and 10 realizations, is shown. The energy density is normalized by the value at the origin. (Small blob, Re$_\lambda=60$)
(\textbf{b}) 
The growth rate of the blob's characteristic radius, $\dot{R}_{\rm blob}$, is compared with predictions from Eq.~\ref{eq:gov_eq}, using coefficients $c_0=1.2$ and $\epsilon_0=0.88$. 
(\textbf{c}) Transport and dissipation coefficients are estimated by minimizing the mean squared error of $\dot{R}_{\rm blob}(t)$ between theory and experimental data. 
(\textbf{d}) Time-evolution of the radial turbulent energy distribution $q(r,t)$.  
Normalizing $q(r,t)$ by the average value in the central region ($r \leq 15$~mm) highlights the evolution of the front shape.
Further  rescaling the spatial coordinate $r$ by a characteristic length scale $h(t) \sim (t-t_0)^{\beta}$ as detailed in SI section IV, yields a best collapse for $\beta=0.38$, in agreement with the analytical prediction of $\beta=0.389$. 
(\textbf{e}) Simulations with experiment-inspired initial conditions reproduce front propagation and the dissipation. The simulation uses ensemble-averaged experiment profile of the small blob at $t = 0.85$ s and uses parameters $(c_0, \epsilon_0) = (1.2, 0.88)$. Applying the same analysis from (c) confirms the predicted scaling collapse.
(\textbf{f}) 
The schematic illustrates the flow of the Cell Dynamical System method to simulate the governing equation. Dissipation is applied via onsite map which flows into a stable fixed point at ($\ell^{-1}, q$)=(0, 0), then the nonlinear transport is applied by a 3D isotropic Laplacian.}
\label{fig: fig5}
\end{figure}

Having examined the late-time decay dynamics of the blob, we now turn our attention to the early stages when the blob expands to fill the chamber as depicted by the kymograph in Figure~\ref{fig: fig5}a and Supplementary Movie 1–2.
Capturing the propagation of a turbulent region requires extending Eq.~\ref{eq:turb_dissipation} to include the effects of gradients in turbulent kinetic energy.
Following  Kolmogorov's seminal work and subsequent extensions~\cite{kolmogorov1942equations,barenblatt1983self, barenblattEvolutionTurbulentBurst1987,barenblatt1996scaling,chen1992renormalization}, the propagation and decay of an isolated free turbulent blob can be described by a model for the spatial evolution of turbulent kinetic energy, consisting of a transport term and a dissipation term:
\begin{align}
\label{eq:b-l_model}
    \partial_t q = \nabla \cdot \left(\kappa_q \nabla q \right) - \epsilon
\end{align}
where the effective diffusion coefficient $\kappa_q$ and the energy dissipation rate $\epsilon$ take forms suggested by Kolmogorov similarity hypotheses~\cite{kolmogorov1942equations},
\begin{align}
    \label{eq:kq_epislon}
    \kappa_q = c_0\,\ell\sqrt{q},\qquad \epsilon = \epsilon_0 \frac{q^{3/2}}{\ell}.
\end{align}
Here, $c_0$ and $\epsilon_0$ are dimensionless parameters that moderate the strength of transport and dissipation respectively.
Substituting Eq.~\ref{eq:kq_epislon} into Eq.~\ref{eq:b-l_model} yields a nonlinear diffusion equation,
\begin{equation}
    \partial_t q = \frac{2}{3} c_0 \ell  \nabla^2 q^{3/2}  - \epsilon_0 \frac{q^{3/2}}{\ell},
    \label{eq:gov_eq}
\end{equation}
which we refer to as the Kolmogorov-Barenblatt turbulent energy balance equation.
Here, both the dissipation term, $-\epsilon_0 q^{3/2}/\ell$, and transport term, $\frac{2}{3} c_0\ell\nabla^2 (q^{3/2})$, are nonlinear in $q$ and depend on a dynamically varying length scale $\ell(t)$, making Eq.~\ref{eq:gov_eq} challenging to solve analytically or numerically \cite{duan2019numerical}.

As a preliminary test of Eq.~\ref{eq:gov_eq}, we compare the predicted expansion rate of the turbulent blob with direct experimental measurements.
For this analysis, we use data from a highly localized blob, which permits extended observation of propagation dynamics before boundary effects become significant, despite its lower initial Taylor Reynolds number (60 vs 200 for the larger blob described up to now).
We define the blob radius, $R_{\rm blob}$, through the normalized second moment of the turbulent energy, $R_{\rm blob}^2(t) =  \int q(\mathbf{x}, t) r^2 d\mathbf{x}~/~\int q(\mathbf{x}, t)  d\mathbf{x}$.
The rate of expansion, $\dot{R}_{\rm blob}$, is presented as the black curve in Figure~\ref{fig: fig5}b.
To predict the instantaneous expansion rate from Eq.~\ref{eq:gov_eq}, we compute the change in energy $\delta q$ over a short interval $\delta t$ from its right-hand side and then deduce $\delta R$ accordingly (see Supplementary Information §5B). 
The evolution of the second moment is governed by both the dissipative term, which nonlinearly decreases the energy, and the diffusive term, which spatially redistributes it.
Figure~\ref{fig: fig5}a illustrates the individual contributions of these terms for a specific pair of values for $c_0$ and $\epsilon_0$; varying these parameters scales each contribution by a constant factor.

We can estimate $c_0$ and $\epsilon_0$ from this instantaneous fit with experiment by minimizing the mean squared error between the predicted total expansion rate and the measured rate averaged over the expansion period ($t=1-5.5$ s), as a function of these parameters. 
Figure~\ref{fig: fig5}c displays an error map in the $c_0$-$\epsilon_0$ plane, revealing a distinct linear valley of minimal error where values of $c_0$ and $\epsilon_0$ are of order unity.
$\epsilon_0$ can also be experimentally estimated using the late-time decay dynamics where $q \sim t^{-2}$ and $\ell$ remains constant at $\ell_{\text{sat}}$. 
By solving $ dq/dt = - \epsilon_0 q^{3/2} / \ell_{\rm sat} $, we obtain $ \epsilon_0 = 2 \ell_{\text{sat}} / [\sqrt{q} (t - t_0)] $, where $\ell_{\rm sat} = 0.26 L_{\rm box}$ for the turbulent blob and $\ell_{\rm sat} = 0.25 L_{\rm box}$ for the double oscillating grid. 
The estimated value of $\epsilon_0$ is $0.88 \pm 0.10$ for the blob, and $1.20 \pm 0.11$ for the double oscillating grid. 
Using this result along with Figure~\ref{fig: fig5}d, we estimate $c_0$ to be $1.2$ during the expansion phase of the blob. 
We confirm that the reconstructed $\dot{R}_{\rm blob}(t)$ is in good agreement with the experiment, as shown by the red curve in Figure~\ref{fig: fig5}b.
Our data provide strong evidence that Eq.~\ref{eq:gov_eq} can effectively capture the spread of turbulence, particularly the predicted nonlinear diffusion and decay, solely based on Kolmogorov similarity hypotheses.

Moving beyond the instantaneous dynamics, we extend the analysis to the evolution of the spatial structure, long time decay and asymptotic scalings. 
To do so, we solve Eq.~\ref{eq:gov_eq} over a prolonged period, with initial conditions derived from the experimental data and using the measured $\ell(t)$.  The dynamics of fronts in this category of equations (related to the porous medium equation) is challenging to capture numerically due to the singular behavior of the nonlinearities \cite{chewSolvingOneDimensionalPorous2021}, and the time-varying coefficients.  
To address this challenge, we develop a split-step numerical approach, using an exact solution of the onsite dynamics, which is then coupled to diffusion.  
This approach is sometimes referred to as the Cell Dynamical System (CDS) method, and has been used to efficiently and accurately simulate other non-equilibrium growth and pattern formation processes ranging from the scaling laws encountered in the kinetics of phase separation \cite{oono1987computationally,shinozaki1993spinodal}, superfluid \cite{mondello1990scaling,mondello1992scaling} and superconducting transitions \cite{liu1991kinetics}, liquid crystal ordering \cite{zapotocky1995kinetics}, as well as structure formation during flow-depositional processes at geothermal hot springs \cite{veysey2008watching,goldenfeld2006dynamics}.  In the latter example \cite{goldenfeld2006dynamics}, it was possible to explicitly compare the solution accuracy of this method with standard numerical methods, going beyond the extraction of scaling exponents.  Unlike direct PDE discretization, the CDS method treats space as a lattice of discrete cells and the time evolution is simulated through injective discrete maps that have fixed points representing the phases of the system.  In our case, this onsite map is the exact solution of the homogeneous decay of $q(\mathbf{x}, t)$.
In our implementation, a single temporal step for $q(\mathbf{x}, t)$ proceeds in two stages, as illustrated in Figure~\ref{fig: fig5}f: first, the local dissipation dynamics is updated via an onsite map derived from an exact integration; subsequently, contributions from the transport term are incorporated by a coupling map for neighboring cells, where the weights in this map are meticulously chosen to preserve the isotropic symmetry of the Laplacian~\cite{thampi2013isotropic}. 
We validate the accuracy of this method in a special case where the governing equation admits an analytical solution (see Supplementary Information \S2C), in particular showing that we can capture not just scaling exponents but the accurate solution of the governing partial differential equation.

Our CDS method for turbulent decay is able to capture easily spatial variations, and thus permits a direct comparison between simulations of Eq.~\ref{eq:gov_eq} and the experimentally measured evolution of the expanding blob, because we can use the experimentally-measured initial conditions in our solution.
Figure~\ref{fig: fig5}a shows a kymograph of the angular-averaged energy field $q_\Omega (r, t) = (1/4\pi)\int q({\mathbf{x}}, t) \delta(|\textbf{x}| - r) d\Omega$ where $\Omega$ is a solid angle. 
In practice, we obtain this kymograph by taking an azimuthal average of ensemble-averaged turbulent energy $\langle q \rangle_{\rm n} (\mathbf{x}, t)$. 
During the early stage ($t$ = 0.6 - 5.5 s), the blob expands freely until it contacts the chamber wall.
Although energy is quickly dissipated throughout this period --— as reflected by the overall decline in intensity --- the turbulent front still propagates outward in a self-similar manner, as shown in Figure~\ref{fig: fig5}d.  
To highlight the shape and the propagation speed of the front, we normalize the radial energy profile by its central value at each time.
Further rescaling $r\rightarrow r/t^{0.38}$ collapses the profiles onto a single curve, indicating the presence of self-similar dynamics during this regime.

To compare with Eq.~\ref{eq:gov_eq} we performed a CDS simulation initialized with the experimental turbulent energy distribution at early time and parametrized by experimentally measured $\ell(t)\sim t^{\gamma}$ with $\gamma=0.36$ for small blobs.
We adopt the parameter combination $c_0=1.2$ and $\epsilon_0=0.88$, identified from the decay of a large blob and the propagation speed of a small blob.
As illustrated in Figure~\ref{fig: fig5}e, the CDS simulation captures both the outward propagation of the energy front and the dynamic scaling collapse observed in experiments, $r \sim t^{0.38}$. The scaling exponent observed in the simulation closely aligns with the theoretical prediction of 0.389.
(see Supplementary Information \S2D).

Previous theoretical studies of Eq.~\ref{eq:gov_eq} primarily focused on the special case where $\ell(t) = \alpha h(t)$, with a dimensionless constant  $\alpha$~\cite{barenblatt1983self, barenblattEvolutionTurbulentBurst1987, chen1992renormalization}. This assumption implies that the size of eddies instantly adapts to the evolving turbulent-non-turbulent interface, represented by $h(t)$. Such a closure gives rise to a self-similar solution in the intermediate asymptotic regime.
The self-similar solution features a sharp front propagating through nonlinear diffusion, with power-law scaling in both front position and energy decay.
Under this assumption of the instantaneous eddy adaptation, the dynamic scaling $\ell(t) \propto h(t) \propto t^{\theta}$ yields an exponent in the range of $0.3-0.4$.

However, our direct measurement of $\ell(t)$ shows such an assumption does not hold generally, and enables a more direct comparison with the general governing equation~\ref{eq:gov_eq}. 
As shown in Figure~\ref{fig: fig2}b,  our measurements of $\ell(t)$ suggests an alternative Ansatz: $\ell(t)\sim t^\gamma$, where $\gamma$ is a constant that reflects specific experimental conditions. We find that its value can be as low as $0.16$ for a large blob.
Although the self-similarity is no longer strictly preserved, the asymptotic solution approaches an approximately self-similar regime, retaining essential characteristics of turbulence propagation, including a sharp front and nonlinear diffusion (see Supplementary Information \S2D).

After examining the spatiotemporal evolution of the blob during the initial free expansion, we now return to the energy decay $\l q \r_{\x, \n}(t)$ over longer timescales.
Figure 2c compares the CDS simulation, using the parameters, estimated from the large blob ($c_0=1.2$, $\epsilon_0=0.88$), with a partially absorbing boundary condition ($15\%$ loss) accommodating some small energy loss at the chamber wall.
While the curve captures the long-time decay, the transition from the initial phase to the $q\sim t^{-2}$ occurs much earlier than observed experimentally (blue curve in Figure~\ref{fig: fig2}c).
Although the exact onset of $q\sim t^{-2}$ depends on boundary conditions, this alone does not account for this discrepancy (see Supplementary Information \S2D).

To investigate the origins for this discrepancy, we examine the 3D turbulent energy field beyond its radial profile. 
In the CDS simulation, the usage of the radial profile ignores the variation in the angular direction, and the Laplacian term $\nabla^2 q^{3/2}$ in Eq.~\ref{eq:gov_eq} rapidly smooths the turbulent energy field even when more spatial variation is accounted for.
In contrast, the radial profile from the experiment shows sustained spatial inhomogeneity near the boundary over extended amount of time. A more detailed inspection of the experimental 2D and 3D datasets for the energy and vorticity fields reveals that eddy-like structures indeed persist throughout the decay (see Supplementary Information \S5A). 
Such persistent eddies may arise from vortex interactions in the bulk and turbulence–boundary interactions~\cite{lighthill1963introduction}, which are physical effects that lie beyond the scope of our mean-field model designed for a freely expanding blob.
To account for these persisting eddy structures without adding further equations to capture these flow effects, we made a minimal modification of Eq.~\ref{eq:gov_eq} by reducing the dimensionless transport parameter $c_0$ in late time, thereby effectively slowing down the homogenization of the flow.
As illustrated by the red curve ($c_0=0.001$) in Figure~\ref{fig: fig2}c, this adjustment indeed delays the transition to the final regime.
Here, both simulations (blue and red curves) use the same experimentally measured $\ell(t)$, indicating that the crossover of decay rate depends not merely on the timing of $\ell(t)$ saturation but also on the detailed spatial structure $q(\mathbf{x}, t)$.
By interpreting $c_0$ as a factor controlling effective transport, reducing it accommodates the persistent eddy-like features and yields better agreement with the delayed transition observed in the experiment.

In conclusion, by measuring the decay of an ensemble of turbulent blobs as well as space-filling turbulence, over several decades in time,  {\it all within the same flow chamber}, we reveal how turbulence decays and propagates. 
Starting from a localized turbulent blob, with an integral length scale that is smaller than the flow chamber, the integral length scale grows in time, and influences the spatiotemporal dynamics of the kinetic energy, in a way that is easily incorporated into our theoretical model and analysis of the data.  When the integral length scale remains constant and the turbulence fully occupies the chamber, a universal $t^{-2}$ decay is observed. Through a combination of data analysis, theoretical and computational modeling, we show how nonlinear diffusion gives rise to sharp fronts separating turbulent from quiescent regions of the flow, and determine experimentally the scaling laws governing propagation and decay that are consistent with theoretical predictions.

\begin{acknowledgments}
Two of us (N.G. and W.T.M.I.) wish to acknowledge  inspiration and discussion (W.T.M.I. and N.G.) and collaboration (N.G.) on the topic of propagation of turbulence with the late Russell Donnelly, in whose lab pioneering observations were made \cite{smithStudyHomogeneousTurbulence1994}, and to personally dedicate this work to him.  This work was partially supported by the U.S. Army
Research Office through Grant No. \#W911NF-17-S-0002, \#W911NF-18-1-0046, and \#W911NF-20-1-0117 to W.T.M.I., and partially supported by a grant to NG from the Simons Foundation (Grant No. 662985, N.G.).
Additional support was provided to WI by the Brown Foundation. 
T.M. acknowledges partial support from Schmidt Science Fellows, in partnership with the Rhodes Trust. 
We also thank LaVision Inc.~for their support on the particle image velocimetry and the particle tracking velocimetry.
The Chicago MRSEC is gratefully acknowledged for access to its shared experimental facilities (US NSF grant DMR2011854). 
For access to computational resources, we thank the University of Chicago’s Research Computing Center and the University of Chicago’s GPU-based high-performance computing system (NSF DMR-1828629).
\end{acknowledgments}

\section*{Author contributions}
W.T.M.I. and N.G. designed project. 
T.M. designed, constructed and performed experiments and experimental data analysis.
N.G and M.Z. devised the theoretical models and simulations. M.Z. wrote and performed CDS simulations and analytical calculations.
All authors contributed to the interpretation of the data and writing the paper.

\section*{Data availability}
The data contained in the plots in this article and other findings of this study are available from the corresponding author on reasonable request.

\section*{Code availability}
The codes to handle 2D PIV to compute energy spectra, structure functions, and dissipation from velocity fields, and to visualize flows are available from the corresponding author upon reasonable request.

\newpage
\onecolumngrid

\newpage

\clearpage

\bibliography{lit.bib}

\newpage
\onecolumngrid
\foreach \i in {1,...,58} {
  \includepdf[pages=\i, pagecommand={\thispagestyle{empty}}]{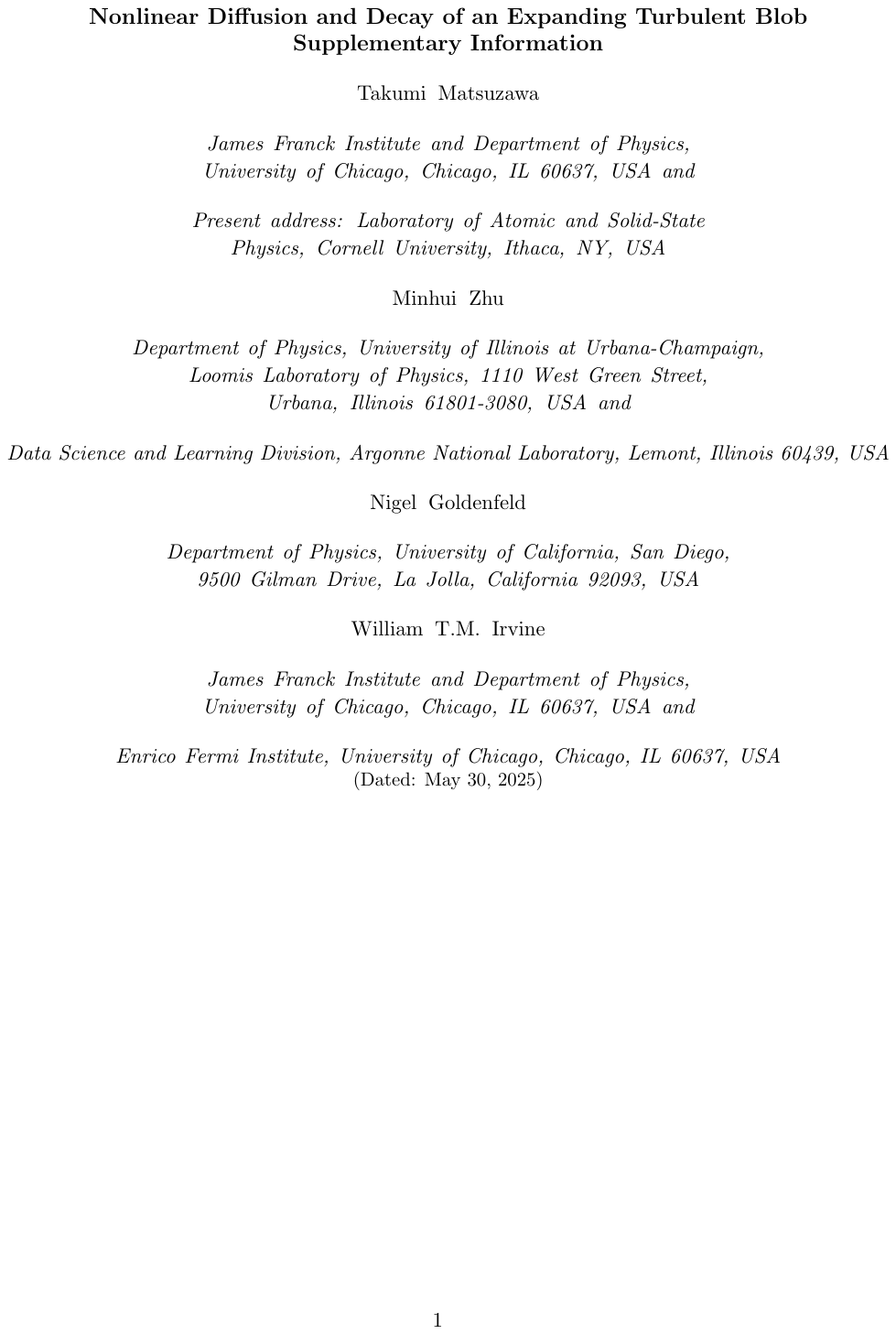}
}

\end{document}